
\magnification \magstep1
\openup 2\jot
\rightline {\sevenrm {BROWN-HET-973}}
\rightline {\sevenrm {DAMTP-94-100}}
\rightline {\sevenrm {December 1994}}
\bigbreak
\leftline {\bf NUCLEOSYNTHESIS CONSTRAINTS ON DEFECT-MEDIATED}
\leftline {\bf ELECTROWEAK BARYOGENESIS}
\medbreak
\leftline {Robert Brandenberger$^{1,2}$, Anne-Christine Davis$^{1,3}$
 and
Martin J. Rees$^{4}$}
\medbreak
\item {$^{1}$} Issac Newton Institute for Mathematical Sciences,
\item {} University of Cambridge, Cambridge CB3 0EH, UK.

\item {$^{2}$} Physics Department, Brown University, Providence RI
 02912, USA.

\item {$^{3}$} Department of Applied Mathematics \& Theoretical Physics
 and
Kings' College,
\item {} University of Cambridge, Cambridge, CB3 9EW, UK

\item {$^{4}$} Institute of Astronomy, University of Cambridge, Cambridge
CB3 OHA, UK.
\vskip 2 truein
\leftline {\bf Abstract}
\medbreak
\noindent
In the defect-mediated electroweak baryogenesis scenario, baryons are
 produced in well separated regions of space. It is shown that between
 the electroweak phase transition at a temperature of $T \sim 100 {\rm
 GeV}$ and the end of nucleosynthesis at $T \sim 1 {\rm KeV}$ the
 baryon inhomogeneities dissipate, and that no constraints on
 defect-mediated electroweak baryogenesis can be derived from
 considerations of inhomogeneous nucleosynthesis.

\vfill \eject
\leftline {\bf 1. Introduction}
\medbreak
\noindent
Recently there has been a lot of interest in the possibility that the
 observed baryon to entropy ratio was generated at the electroweak
 scale [1-3] (for recent reviews see e.g. Refs. 4 and 5).  Most
 electroweak baryogenesis models assume that the electroweak phase
 transition was first order and proceeded via the nucleation, expansion
 and subsequent percolation of critical bubbles. In this case, the
 baryon distribution after the bubble percolation is essentially
 homogeneous since the regions of net baryon production, the bubble
 walls, sweep out all of space.

The order of the electroweak phase transition, however, is not known.
 Even if the transition is first order, its dynamics may be driven by
 thermal fluctuations. Recent simulations in scalar field theories [6]
 and evidence from condensed matter systems (see e.g. Ref. 7 and
 references therein) argue against nucleation of critical bubbles as
 the mechanism driving the phase transition.

In Refs, 8 and 9, a new mechanism of baryogenesis was proposed which
 is independent of the order and detailed dynamics of the electroweak
 phase transition. In this theory, baryon production is mediated by
 topological defects produced at an energy scale $\eta$ equal or
 higher than the scale $\eta_{EW}$ of electroweak symmetry breaking.
 Provided that the electroweak symmetry is restored in the core of the
 defects, baryons are produced there during the out-of-equilibrium
 motion of the defects. In this scenario, baryons are produced
 inhomogeneously.

There are severe constraints on inhomogeneous nucleosynthesis. The
 success of homogeneous big bang nucleosynthesis in explaining the
 observed abundances of light elements[10] (in particular
 $^{4}He,~d + ^{3}He$ and $^{7}Li$) makes it hard to allow for any
 inhomogeneities in the baryon distribution at the time when
 nucleosynthesis begins. This leads to possible constraints on
 scenarios with inhomogeneous baryogenesis [11,12].

Consider this issue in more detail. The predictions of homogeneous big
 bang nucleosynthesis for the light elements $^{4}He,~d + ^{3}He$ and
 $^{7}Li$ are compatible with the observed abundances only in a narrow
 interval of the baryon to entropy ratio $\hat {\eta} =
 n_{b}/n_{\gamma}$ $(n_{b}$ and $n_{\gamma}$ are the baryon and photon
 number densities, respectively):
$$
3 \times 10^{-10} < {\hat {\eta}} < 10^{-9} \ \ \ . \eqno (1)
$$
Any significantly higher or lower values of $\hat {\eta}$ lead to an
 overproduction of $^{7}Li$. A lower value of $\hat {\eta}$ leads to
 an overproduction of $d + ^{3}He$ but to a deficit of $^{4}He$, and a
 higher value of $\hat {\eta}$ induces overproduction of $^{4}He$ but
 a deficit of $d + ^{3}He$. Hence, if at the onset of nucleosynthesis
 the baryons are inhomogeneously distributed, with regions of $0 < \hat
 {\eta} < 3 \times 10^{-10}$ surrounding regions with $\hat {\eta} < 10^{-9}$
 but with the same space-averaged value of $\hat {\eta}$, then an
 overproduction of $^{7}Li$ will result. This issue has recently been
 analysed in detail in Ref. 13 (see Ref. 14 for a selection of early
 references). Note that it is possible to construct special
 inhomogeneous nucleosynthesis scenarios.  For example[15], if
 there are no baryons at all in the ``low baryon density'' regions, then
 agreement with observations can be obtained by lowering the overall
 value of $\hat {\eta}$ such that $\hat {\eta}$ lies in the range (1)
 in the high baryon density regions, thus leading to a low
 $\Omega_{B}$ Universe.

In this letter we investigate whether the defect-mediated electroweak
baryogenesis scenario is constrained by nucleosynthesis
considerations.  In order to answer this question one must study the
dissipation of baryon inhomogeneities between $T \sim 100 {\rm GeV}
= T_{EW}$ (the electroweak phase transition scale) and $T \sim 1
{\rm KeV}$ (the end of nucleosynthesis). Our conclusion is that for
most parameter values of the defect-mediated baryogenesis scenarios
studied, the combined effects of neutrino inflation and baryon
diffusion are sufficiently strong to homogenize the baryon
distribution by the temperature $T \sim 100 \ {\rm KeV}$. Hence, no
constraints can be derived.
\bigbreak
\leftline {\bf 2. Disspation of Baryon Inhomogeneities}
\medbreak
\noindent
The dissipation of baryon inhomogeneities at temperatures between 100
 GeV and 1 KeV has recently been studied in great detail in Refs. 16
 \& 17 (see also Ref. 12). The most important processes are baryon
 diffusion and neutrino diffusion.

At temperatures above about 1 MeV, neutrinos are in thermal
equilibrium with the plasma. Since they are the particles with the longest
mean free path, they are the most efficient ones at transporting
energy.

In an inhomogeneous electroweak baryogenesis model, entropy
perturbations are produced during the electroweak phase
transitions. By the equation of pressure equilibrium, regions with an
overdensity of baryons must have a lower than average
temperature. Neutrino heating of such a cold baryon-rich region will
cause it to expand in order to maintain pressure
equilibrium [16,17]. This effect is called``neutrino inflation''.

Neutrino inflation has the effect of lowering the amplitude of the
inhomogeneities, but not washing them out entirely. The key length
scale in electroweak baryogenesis is the Hubble radius at $T_{EW}$
which is
$$
t_{EW} \simeq {1 \over (10g^{*})^{1/2}} m_{pl} \eta^{-2}_{EW} \simeq
 0.3 cm
\ \ \ . \eqno (2)
$$
Here, $g^{\ast}$ is the number of spin degrees of freedom of the thermal
 bath, and $m_{pl}$ is the Planck mass. If $\lambda$ is the diameter
 of a region with baryon excess, then neutrino inflation will between
 100 GeV and 1MeV reduce the amplitude of such a baryon perturbation
 to a value $A$ which depends on $\lambda [16,~17]$

$$\eqalignno{
A & \simeq 10^{4} ~{\rm for}~~\lambda \, \epsilon \, [10^{-7},~10^{-1}
 ] {\rm
cm}\cr
A & \simeq \left ( {\lambda \over 10^{-15}} \right )^{1/2} ~{\rm
 for}~
\lambda \, \epsilon \, [ 10^{-15},~ 10^{-7} ] {\rm cm} & (3)\cr
A & = 1 ~{\rm for}~ \lambda < 10^{-15} {\rm cm}\cr
}$$
In the above, $\lambda$ is the physical size at $T_{EW}$. For
defect-mediated baryogenesis, only values of $\lambda$ smaller than
$t_{EW}$ are of interest. We have implicitly assumed that the initial
value of $A$ exceeds the value given in (3).

After the neutrinos fall out of thermal equilibrium, neutrino inflation
ceases to be an effective energy transport mechanism. It has been
 shown
[17] that for $T < 1 {\rm MeV}$, baryon diffusion is the dominant way of
dissipating entropy fluctuations. The baryon diffusion length $l_{diff}$
depends on the initial value of $A$ (at $1 {\rm MeV}$). An approximate
expression for
$l_{diff}$ is [17]
$$
l_{diff} \simeq 0.1 \left ( {A \over A_{0}} \right )^{1/2} {\rm cm},~~~
 A
\geq A_{0} \ \ \ , \eqno (4)
$$
with $A_{0} = 10^{2}$ (note that $l_{diff}$ is expressed in terms of
 physical size at $100 \ {\rm GeV}$ of a given comoving scale). Baryon
 inhomogeneities on scales smaller than $l_{diff}$ get erased by
 diffusion. As is evident from Figs. 14 - 16 of Ref. 17, the
 distribution of baryons has become essentially homogeneous already at
 $T= 0.1 \ {\rm MeV}$, the onset of nucleosynthesis.

To summarize, the evolution of baryon inhomogeneities produced during
 the electroweak phase transition proceeds in two stages. Between 100
 GeV and 1 MeV, the amplitude of the perturbations decreases by
 neutrino diffusion.  Below 1 MeV, baryon diffusion becomes dominant
 and spreads out the baryons.

In order to study possible constraints on defect-mediated baryogenesis
 from nucleosynthesis considerations, we must know the initial
 amplitude $A$ of the entropy perturbations, and the mean separation
 $d$ of the defects inducing baryogenesis. If
$$
d < l_{diff} (A) \ \ \ , \eqno (5)
$$
then the baryons have homogenized by the time of the onset of
nucleosynthesis, and the scenario is safe.

In this paper we will assume that independent of the scale $\eta$ at
 which the defects are produced and independent of the type of defects
 considered, the mean baryon to entropy ratio $\hat {\eta}$ lies in
 the interval (1).
\bigbreak
\leftline {\bf 3. Constraints on Defect-Mediated Electroweak Baryogenesis}
\medbreak
\noindent
In the defect-mediated electroweak baryogenesis scenario of Refs.  8
 and 9, baryons are produced inside of moving topological defects in
 which the electroweak symmetry is restored. Baryon number violating
 electroweak sphaleron processes are unsuppressed in the defect cores,
 CP violation is enhanced in the defect walls (in models with extra CP
 violation in the Higgs sector such as the two doublet model, a theory
 commonly used to study electroweak baryogenesis), and the defect
 dynamics is out of equilibrium.  Thus, all of Sakharov's
 criteria [18] for baryogenesis are satisfied.

In the baryogenesis scenario of Refs. 8 and 9, baryons are thus only
produced in regions swept out by the topological defects following the
electroweak phase transition. The size of the baryon-rich regions
depends on the spatial extent of the particular type of defect which
is catalyzing baryogenesis, and the mean separation of these regions
is determined by the mean separation of the defects. Thus, we have a
realization of the situation studied in Refs. 16 and 17 in which
baryons are produced inhomogeneously.

If $V_{BG}$ is the volume participating in electroweak baryogenesis,
 and $V$ is the total volume, then the amplitude of the baryon
 perturbation is
$$
A = {V \over V_{BG}} \ \ . \eqno (6)
$$
Knowing this amplitude and the mean separation of the regions where
 baryons are produced, it is possible to verify whether the condition
 (5) is satisfied, i.e. whether the baryon inhomogeneities can
 dissipate sufficiently or not to effect nucleosynthesis.

In the following we shall investigate domain wall and cosmic string
mediated baryogenesis. Domain walls moving with velocity $v_{D}$ in
their normal direction will sweep out a fraction of the order $v_{D}$
of space.  Hence
$$
A \sim v^{-1}_{D} \ \ . \eqno (7)
$$
For domain walls forming at a scale $\eta$ close to $\eta_{EW}$, the
 mean separation $d$ will be microscopic
$$
d \left ( t_{EW} \right ) \sim \eta^{-1} \left ( {\eta \over \eta_{EW}
 }
\right ) ^{p} ,\eqno (8)
$$
where $p$ is some power determined by the evolution of the domain wall
 network between $T= \eta$ and $T=\eta_{EW}$. For domain walls,
 $\eta$ must be close to $\eta_{EW}$, otherwise the Universe would be
 domain wall dominated at $\eta_{EW}$. Hence, generically

$$
d \left ( t_{EW} \right ) \ll l_{diff} (A) \ \ , \eqno (9)
$$
and no additional constraints on the model result from the
considerations of this work. Note that even if domain walls are
produced at $\eta$ close to $\eta_{EW}$, they must disappear by the
present time in order to ensure that the Universe is not dominated by
walls today.

For model building, a theory in which cosmic strings mediate electroweak
baryogenesis is less constrained. The evolution of a network of cosmic
strings has been studied in great detail. For

$$
\eta < (m_{pl} \eta_{EW} )^{1/2} \simeq 3 \times 10^{10} \ {\rm GeV} \eqno (10)
$$
the evolution of the string network is still friction-dominated, for
 larger values of $\eta$ the strings are in their scaling regime.

For friction-dominated strings, the mean separation is [19]

$$
d = \xi (t_{EW}) \sim (G{\mu})^{1/2} m_{pl}^{1/4} t_{EW}^{5/4}
 \eqno (11)
$$
where $\mu \simeq \eta^{2}$ is the mass per unit length of the string.
 The length $\xi (t_{EW})$ also determines the volume in which
 baryogenesis takes place. Thus the initial amplitude $A_{in}$ of the
 baryon perturbation is

$$
A_{in} \sim \left ( {t_{EW} \over \xi (t_{EW})} \right ) ^{2}  {t_{EW}
 \over
R_{s}} \gg 10^{4} ,\eqno (12)
$$
where $R_{s}$ is the radius about the string to which the electroweak
symmetry is restored. Except for values of $\eta$ very close to $\eta_{EW}$
$$
{d \over t_{EW}} > 10^{-7} \ \ . \eqno (13)
$$
Hence, from (3) it follows that after neutrino diffusion, the amplitude of the
 baryon perturbations will be
$$
A \simeq 10^{4} \ \ , \eqno (14)
$$
the value we will use for all strings in the friction era. With the
 above value for  $A$, one can easily evaluate the criterion (5). We
 find that for
$$
\eta < 10^{9} \ \ , \eqno (15)
$$
equation (5) is satisfied.

We thus conclude that for $\eta < 10^{9}$, the baryon distribution at
 the onset of nucleosynthesis in cosmic string-mediated electroweak
 baryogenesis is homogeneous.

For $\eta > 3 \times 10^{10}$, strings are in their scaling regime. In the
scaling regime, the string network consists of long strings with
curvature radius and mean separation proportional to the Hubble
radius, and of a distribution of loops with number density [20]

$$
n (R,t) = \cases{
\nu R^{-2} t^{-2} & $\gamma G \mu t < R < \alpha t$\cr
\nu (\gamma G \mu)^{-2} t^{-4} & $R < \gamma G \mu t = R_{c}$\cr
} \eqno (16)
$$
where $\alpha$, $\gamma$ and $\nu$ are constants which must be
determined in numerical simulations. The constant $\alpha$ is a
measure of the scale $\alpha t$ at which loops are produced at time
$t$. Present simulations[21] indicate $\alpha < 10^{-2}$. The value
of $\nu$ is at present also poorly determined, but is [21] of the
order 1, and $\gamma$ measures the strength of gravitational
radiation ($\gamma \sim 50$ according to Ref. 22).

The separation $d$ between baryogenesis volumes is given by the separation
 of
small loops which is

$$
d \simeq (R_{c} \nu (\gamma G \mu)^{-2} t^{-4} )^{-1/3} = \nu^{-1/3}
(\gamma G \mu)^{1/3} t_{EW} .\eqno (17)
$$
{}From (3) it follows that
$$
A \simeq 10^{4}
$$
(the initial amplitude $A_{in}$ for scaling strings is much larger
 than $A$ because the strings occupy only a small volume of space). We
 can now investigate under which conditions the criterion (5) is
 satisfied.  The result is
$$
G{\mu} < \nu \gamma^{-1} \eqno (19)
$$
or
$${\eta \over m_{pl}} < (\nu \gamma^{-1})^{1/2} \eqno (20)
$$
This implies that for all particle physics motivated cosmic strings in
 the scaling regime, baryons have homogenized by the onset of
 nucleosynthesis.

At first sight it seems that for a narrow range of values of $\eta$
$$
10^{9} {\rm GeV} < \eta < 3.10^{10} \ {\rm GeV} \ \ , \eqno (21)
$$
defect-mediated electroweak baryogenesis is constrained by
nucleosynthesis.  However, for this range of values of $\eta$ there
will already be many loops as part of the string distribution. Hence,
the mean separation of baryogenesis sites will be smaller than that
given in Eq. (11), and the criterion (5) will in fact be satisfied.
\bigbreak
\leftline {\bf 4. Conclusions}
\medbreak
\noindent
We have shown that the combined effects of neutrino inflation and
 baryon diffusion are efficient enough to homogenize the baryon
 distribution in the defect-mediated electroweak baryogenesis
 scenarios we have considered, in spite of the initial large amplitude
 of the baryon perturbations.

Neutrino inflation is crucial in order to obtain this result, since it
 leads to a decrease of the amplitude of entropy inhomogeneities to a
 value $A \simeq 10^{4}$ between $100 \ {\rm GeV}$ and $1\ {\rm MeV}$. With
this value of $A$ the comoving
 baryon diffusion length is comparable to the Hubble radius at
 $t_{EW}$ (when translated to physical length at $t_{EW}$). In cosmic
 string-mediated models of electroweak baryogenesis the mean
 separation of the defects turns out to be much smaller than the
 baryon diffusion length, unless $\eta$ is very close to the Planck
 scale. Note that without taking neutrino diffusion into account, the
 effective baryon dissipation length would have been so small as to
 render most of the models in conflict with nucleosynthesis.

We have shown that the only defect-mediated models of electroweak
baryogenesis which are endangered by the inhomogeneous nucleosynthesis
constraints are theories in which the mean separation of the defects
 at
$t_{EW}$ exceeds about 3\% of the Hubble radius at $t_{EW}$ (see
 Eqs. (4)
\& (5)).
\bigbreak
\leftline {\bf Acknowledgements}
\medbreak
\noindent
This work was supported in part by US Department of Energy under
 grant DE-FG029
\hfill \break1ER40688, Task A, by PPARC and the Royal Society in
 the UK and
by an NSF-SERC collaborative research award NSF-INI-9022895. We thank
 the
Isaac Newton Institute for hospitality.
\vfill \eject
\leftline {\bf References}
\medbreak
\item {1.} M. Shaposhnikov, {\it Nucl. Phys.} {\bf B287}, 757 (1987);
\item {} M. Shaposhnikov, {\it Nucl. Phys.} {\bf B299}, 797 (1988);
\item {} L. McLerran, {\it Phys. Rev. Lett.} {\bf 62}, 1075 (1989).

\item {2.} N. Turok and J. Zadrozny, {\it Phys. Rev. Lett.} {\bf
 65}, 2331
(1990);
\item {}N. Turok and J. Zadrozny, {\it Nucl. Phys.} {\bf B358}, 471
 (1991);
\item {} L. McLerran, M. Shaposhnikov, N. Turok and M. Voloshin {\it
 Phys.
Lett.} {\bf B256}, 451 (1991).

\item {3.} A. Cohen, D. Kaplan and A. Nelson {\it Phys. Lett.} {\bf
 B245},
561 (1990);
\item {} A. Cohen, D. Kaplan aaand A. Nelson {\it Nucl. Phys.} {\bf
 B349},
727 (1991);
\item {} A. Nelson, D. Kaplan and A. Cohen, {\it Nucl. Phys.} {\bf
 B373},
453 (1992).

\item {4.} N. Turok in ``Perspectives on Higgs Physics'', ed. G. Kane
 (World
Scientific, Singapore 1992).

\item {5.} A. Cohen, D. Kaplan and A. Nelson, {\it Ann. Rev. Nucl.
 Pert.
Sci.} {\bf 43}, 27 (1993).

\item {6.} J. Borrill and M. Gleiser, ``Thermal Phase Mixing During
 First
Order Phase Transitions'', Dartmouth preprint DART-HEP-94/06 (1994).

\item {7.} N. Goldenfeld, ``Dynamics of Cosmological Phase Transitions:
What can we learn from Condensed Matter Physics'', Univ. of Illinois
preprint (1994), to be publ. in ``Formation and Interactions of Topological
Defects'', A.-C. Davis and R. Brandenberger (eds.), (Plenum Press,
 New
York, 1995).

\item {8.} R. Brandenberger, A.-C. Davis and M. Trodden, {\it Phys.
 Lett.}
{\bf B335}, 123 (1994).

\item {9.} R. Brandenberger, A.-C. Davis, T. Prokopec and M. Trodden,
``Local and Nonlocal Defect-Mediated Electroweak Baryogenesis'', Brown
preprint BROWN-HET-962 (1994).

\item {10.} for a recent review see e.g. B. Pagel, {\it Ann. N.Y.
 Acad.
Sci.} {\bf 647}, 131 (1991).

\item {11.} G. Fuller, K. Jedamzik, G. Mathews and A. Olinto, {\it
 Phys.
Lett.} {\bf B333}, 135 (1994).

\item {12.} A. Heckler, ``The Effects of Electro-weak Phase-Transition
Dynamics on Baryogenesis and Primordial Nuclosynthesis'', Univ. of
Washington thesis (1994), astro-ph/9407064.

\item {13.} K. Jedamzik, G. Fuller and G. Mathews, {\it Ap. J. } {\bf
 423},
50 (1994).

\item {14.} R. Epstein and V. Petrosian,  {\it Ap. J. } {\bf 197},
 281 (1975);
\item {} J. Applegate, C. Hogan and R. Scherrer, {\it Phys. Rev.}
 {\bf
D35}, 1151 (1987);
\item {} C. Alcock, G. Fuller and G. Mathews, {\it Ap. J. } {\bf 320},
 439
(1987);
\item {} R. Malaney and W. Fowler, {\it Ap. J. } {\bf 333}, 14 (1988);
\item {} H. Kurki-Suonio et al. {\it Phys. Rev.} {\bf D38}, 1091
 (1988);
\item {} N. Teraswawa and K. Sato, {\it Prog. Theor. Phys.} {\bf
 81}, 254
(1989);
\item {} R. Malaney and G. Mathews, {\it Phys. Rep.} {\bf 229}, 145
 (1993).

\item {15.} K. Jedamzik, G. Mathews and G. Fuller, ``Absence of a
 lower
limit on $\Omega_{b}$ in Inhomogeneous Primordial Nucleosynthesis'', {\it
Astro-Ph.} 9407035 (1994).

\item {16.} A. Heckler and C. Hogan, {\it Phys. Rev.} {\bf D47},
 4256 (1993).

\item {17.} K. Jedamzik and G. Fuller {\it Ap. J. } {\bf 423}, 33
 (1994).

\item {18.} A. Sakharov, {\it Pisma Zh. Eksp. Teor. Fiz} {\bf 5},
 32 (1967).

\item {19.}T.W.B. Kibble, {\it Acta Phys. Pol.} {\bf B13}, 723 (1982);
\item {} A. Everett, {\it Phys. Rev.} {\bf D24}, 858 (1981);
\item {} M. Hindmarsh, Ph.D thesis, University of London (unpublished,
 1986).

\item {20.} N. Turok and R. Brandenberger, {\it Phys. Rev.} {\bf
 D33},
2175 (1986);
\item {} A. Stebbins, {\it Ap. J. (Lett.) } {\bf 303}, L21 (1986);
\item {} H. Sato, {\it Prog. Theor. Phys.} {\bf 75}, 1342 (1986).

\item {21.} D. Bennett and F. Bouchet, {\it Phys. Rev. Lett.} {\bf
 60}, 257
(1988);
\item {} B. Allen and E.P.S. Shellard, {\it Phys. Rev. Lett.} {\bf
 64}, 119
(1990);
\item {} A. Albrecht and N. Turok, {\it Phys. Rev.} {\bf D40}, 973
 (1989).

\item {22.} T. Vachaspati and A. Vilenkin {\it Phys. Rev.} {\bf D31},
 3052
(1985);
\item {} N. Turok {\it Nucl. Phys.} {\bf B242}, 520 (1984)
\item {} C. Burden {\it Phys. Lett.} {\bf 164B}, 277 (1985).

\vfill \eject

\bye